\documentclass{article}
\usepackage{epsf}
\usepackage{emulateapj}
\input{psfig}
\newcommand{\etal}{{\it et~al. }}

\begin{document}

\begin{center}
\title{ROSAT Observations of the Vela Pulsar}

\author{F.D. Seward\altaffilmark{1},
M. A. Alpar\altaffilmark{2},
C. Flanagan\altaffilmark{3},
 \"{U}. Kizilo\v{g}lu\altaffilmark{2},
C. Markwardt\altaffilmark{4},
P. McCulloch\altaffilmark{5},
H. \"{O}gelman\altaffilmark{6}}

\end{center}

\altaffiltext{1}{SAO, 60 Garden Street, Cambridge, MA  02138}
\altaffiltext{2}{Middle East Technical University, Astrophysics
Div., Inonu Blvd., 06531 Ankara, Turkey}
\altaffiltext{3}{Hartebeesthoek Radio Observatory, P.O. Box 443,
Krugersdorp, Transvaal 1740, South Africa}
\altaffiltext{4}{NASA/GSFC, Code 662, Greenbelt, MD 20771}
\altaffiltext{5}{University of Tasmania, School of Physics, GPO Box
252-21, Hobart, Tasmania, Australia} 
\altaffiltext{6}{University of Wisconsin, Department of Physics, 1150
University Avenue, Madison, WI 53706}

\slugcomment{To appear in The Astrophysical Journal}

\begin{abstract}
\noindent

The ROSAT HRI was used to monitor X-ray emission from the Vela Pulsar.
Six observations span 2-1/2 years and 3 glitches.  The summed data
yield a determination of the pulse shape, and X-ray emission from
the pulsar is found to be 12\% pulsed with one broad and two narrow
peaks.  One observation occurred 15 days after a large glitch.  No
change in pulse structure was observed and any change in X-ray
luminosity, if present, was less than 3\%.  Implications for neutron
star structure are discussed. \\
\end{abstract}


\section{INTRODUCTION}

The Vela Pulsar, located at the center of
a prominent supernova remnant and with characteristic age ($p/2\dot{p}$)
of $1.1\times 10^{4}$ years, is ``young''.  The distance to the pulsar is
only 500 pc (Frail and Weisberg, 1990) and the column density of 
absorbing material ($\approx 2 \times 10^{20}$ atoms cm$^{-2}$ or less) is
low.  Thus, soft x-rays from the pulsar can be detected and have been
interpreted as thermal radiation from the surface of the hot neutron
star. (Harnden \etal 1985, \"{O}gelman \etal 1993). \\

Like the younger Crab Pulsar, the Vela Pulsar is a source of high
energy gamma-rays (Kanbach et al 1994).  It is well known that the
Crab Pulsar exhibits approximately
the same pulse structure at $\gamma$-ray, X-ray, and optical
wavelengths, two rather narrow pulses, separated by $\approx$ 0.4 in
phase.  The Vela Pulsar, however, emits a pulse waveform which varies with
energy (Grenier \etal 1988).  At 100 MeV, there are two narrow pulses 
with width and spacing almost
identical to those from the Crab Pulsar.  In visible light, there are
two broad pulses separated by only $\approx$ 0.2 in
phase.  X-ray pulsations from the Vela Pulsar were recently
discovered by \"{O}gelman et al (1993).  They found the pulsed fraction 
to be $\approx$
10\%  (in contrast to $\approx$ 100 \% for the Crab Pulsar) and the
pulse structure, complex. \\

Like other isolated pulsars, the period of the Vela Pulsar, has been
observed to increase steadily and regularly with time.  There are,
however, in the Vela Pulsar, discontinuities  or ``glitches'' in the
timing which occur on average every 2.5 years.  To observe this
phenomenon, the radio signal is
monitored daily (McCulloch et al 1988).  During a glitch, the
frequency of radio pulsations,  $\nu$, increases suddenly by
$\approx 1$ part in $10^{6}$ and the frequency derivative, $\dot \nu$,
increases by $\approx$ 1 part in 10$^{2}$ and then recovers steadily
from this transient behavior.  Duration of the recovery is 10 - 100 days. \\

Glitches offer a rare opportunity to learn something
about the interior of the neutron star and much has been learned from
radio-timing data alone (Link et al 1992).  It is believed that,
during the glitch, rotational energy is dissipated in the stellar
interior.  The time scale of energy flow depends on the size of the
star, the nature of the glitch mechanism, and the structure of the
outer layers of the star.  The amount of
thermal energy generated and conducted to the surface could 
be enough to cause an observable increase in temperature of the
surface (van Riper et al 1991).  We attempted to
observe this with ROSAT. \\

\begin{center}
\section{ROSAT OBSERVATIONS}
\end{center}

The requirement that the ROSAT solar panels point at the sun $\pm
15^{\circ}$ forbid observations of the Vela Pulsar except
during the intervals, 22 April to 26 June and 25
October to 26 December.  During these intervals, we obtained data with
the ROSAT HRI 6 times.  There were also two previous ROSAT PSPC
observations.  Parameters of the observations are
listed in Tables 1 and 2, and exposure times range from 20 to 60 ks..  Figure
1 compares the dates of ROSAT observations with times of recent glitches. \\

The time structure of the ROSAT observations was not ideal for pulsar work.
Typically the target was observed for 1000 - 3000 seconds during each
95-minute orbit but with some long intervals devoted to other targets.
A typical data stream for an observation contains $\sim$ 10
hours exposure spread over $\sim$ 15 days time. \\

\begin{center}
\psfig{figure=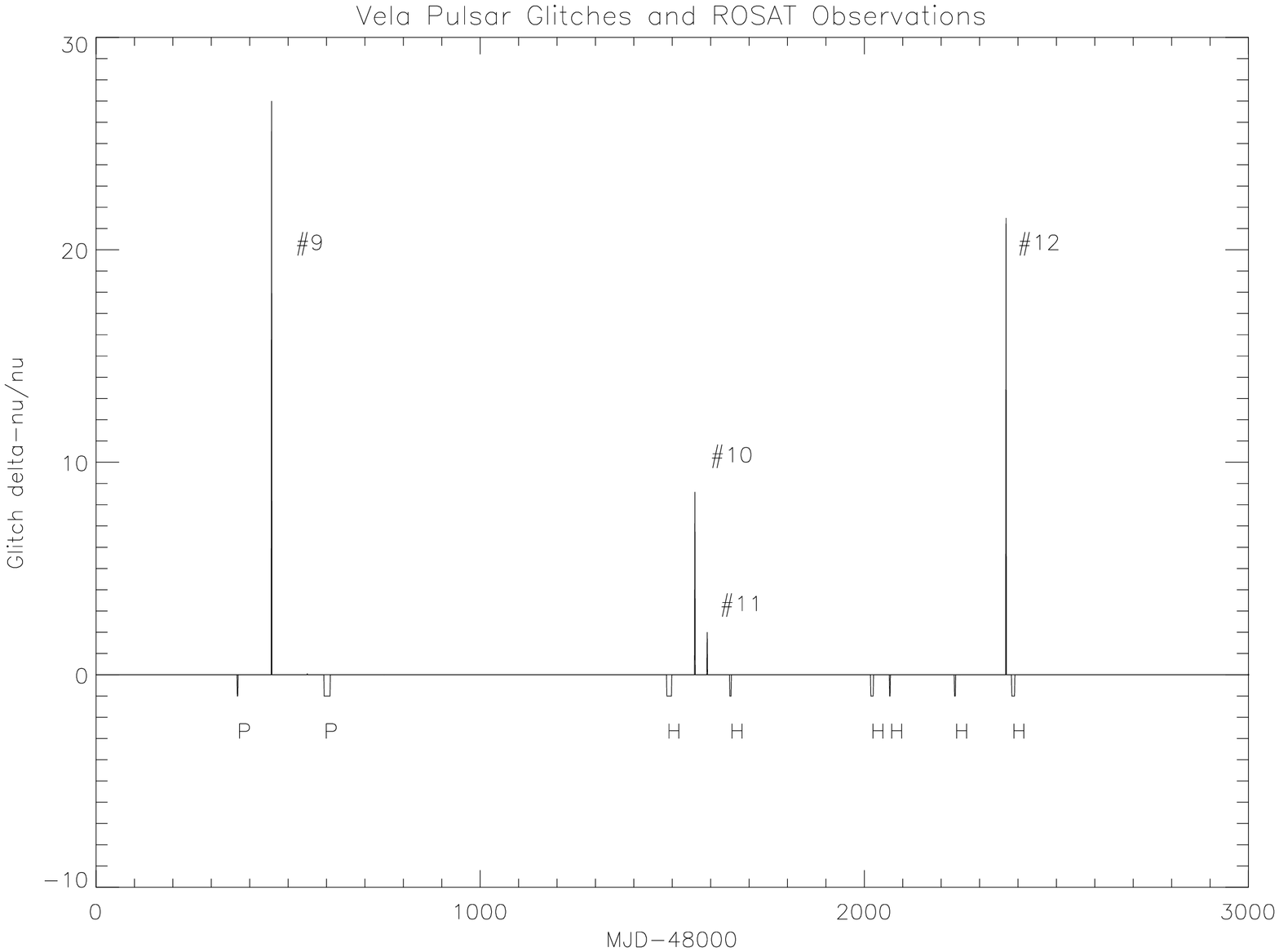,angle=90,width=3.5in}
\begin{minipage}[h]{3.5in}
{\small
Figure 1:  Time history of Vela Pulsar glitches and ROSAT observations. 
Observations are below the horizontal line. P indicates 
PSPC, H indicates HRI.  Width of the pip
shows time spread of data during that observation.  Glitches are above the line.
Height of the glitch-line is proportional to $\Delta\nu/\nu$
} 
\end{minipage}
\end{center}

In the longest exposure (60 ks spread over 14 days) the pulsed signal
was strong enough so the pulsar frequency could have been identified
using x-ray data alone, even without the known radio-measured pulsar 
frequency.  In the
shortest exposure, it was necessary to know the approximate 
radio-measured pulsar frequency to find the x-ray period. \\

We searched for the most significant x-ray period by epoch folding
over a range bracketing the period calculated from a radio period
measured within a few days of the ROSAT observation.  In every case,
the period calculated from the radio ephemeris agreed within a
few ns with the most significant x-ray period in that range.  The
derived X-ray pulse shape, however, is very dependent on the value 
used for the period.  A shift of a few ns makes a 
difference.  We therefore used the Hartebeesthoek and the U. Tasmania
radio data to calculate the
expected period to 0.1 ns and used this period to fold the x-ray
data.   Table 1 lists the input radio data
and table 2, the pulse periods calculated (and observed) at the start
of each ROSAT observation. \\

We tried to use the radio observations to calculate the absolute phase
of x-ray data but got errors of $\sim$ 0.2 in phase.  To better
determine the fine structure of 
the x-ray pulse, we aligned the HRI observation in phase using the
highest peak and summed the observations.  40 phase bins were used.
If this procedure were 
followed using random data, we would, of course, get a single, apparently
significant, peak in the summed light curve, but there should not be
other significant structure.  Since the summed Vela Pulsar light curve
shows a second sharp peak, we believe this procedure is valid and
the observed structure to be real. \\

Figure 2 shows the light curve generated by adding the 3 longest
pre-Oct-96-glitch observations (May 94, Oct 94, June 96).
  The pulse structure was clear in each of
these and the phase alignment was done to an accuracy of .025 in
phase.  The same shape is apparent in all the
longer individual HRI observations and, is similar to that
reported by \"{O}gelman \etal (1993) for the 1991 PSPC observations.
There are 2, sometimes 3, narrow peaks superimposed on a broader
structure.  The Oct 95 and Dec 95 light curves were not as good
statistically and the structure not as clear.  The appearance of
shorter observations is sometimes almost a sawtooth with one prominent
peak and valley. \\

\begin{center}
\psfig{figure=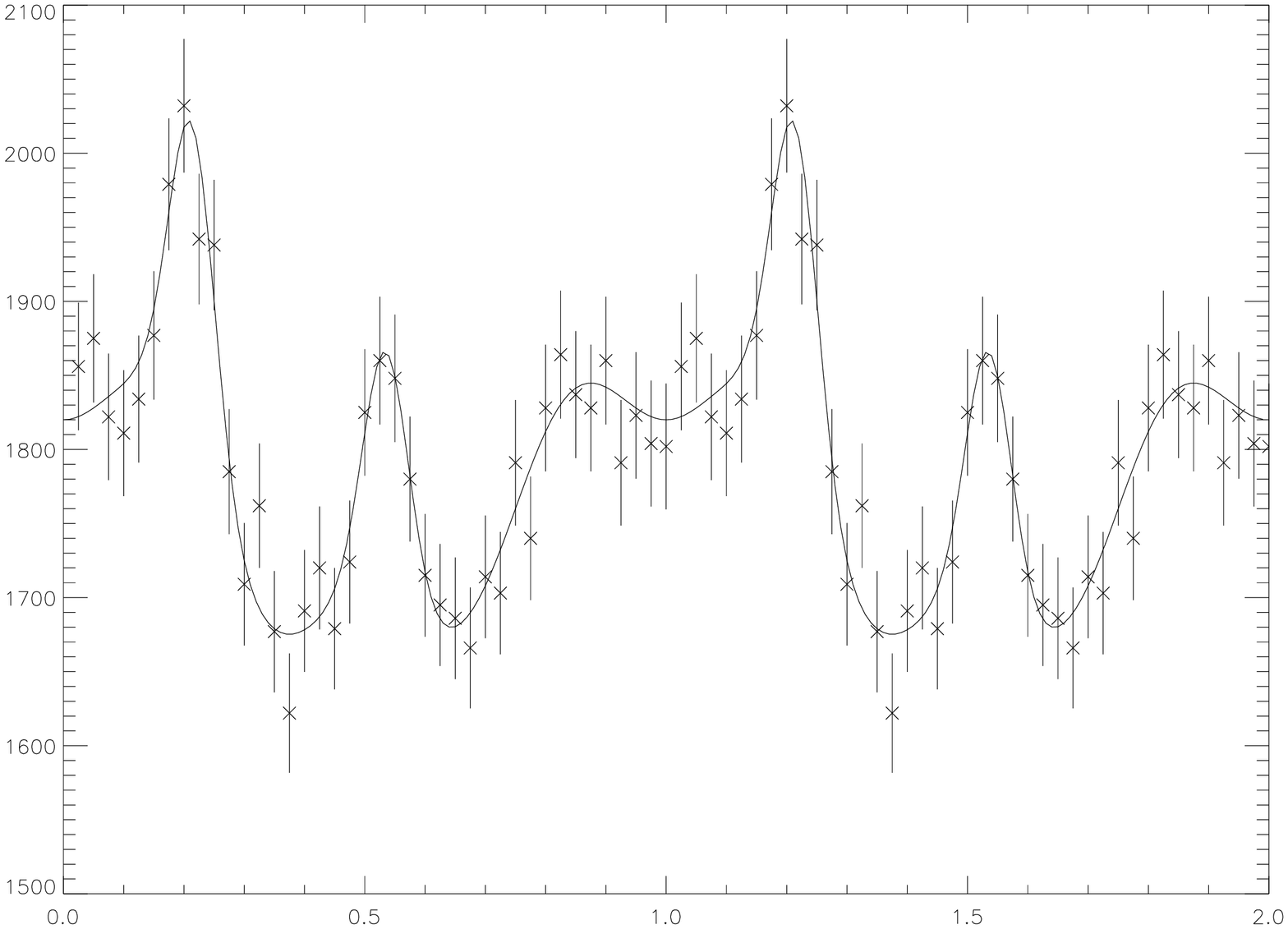,angle=90,width=3.5in}
\begin{minipage}[h]{3.5in}
{\small
Figure 2:  Pre-glitch Vela Pulsar light curve.  HRI count rate vs. phase 
over two complete cycles.  The 3 longest pre-glitch HRI
observations have been phase-matched using the strongest peak,
summed, and fit with an empirical curve.  Errors shown are one
sigma.  Total observation time was 123 ks.
}
\end{minipage}
\end{center}

\begin{center}
\psfig{figure=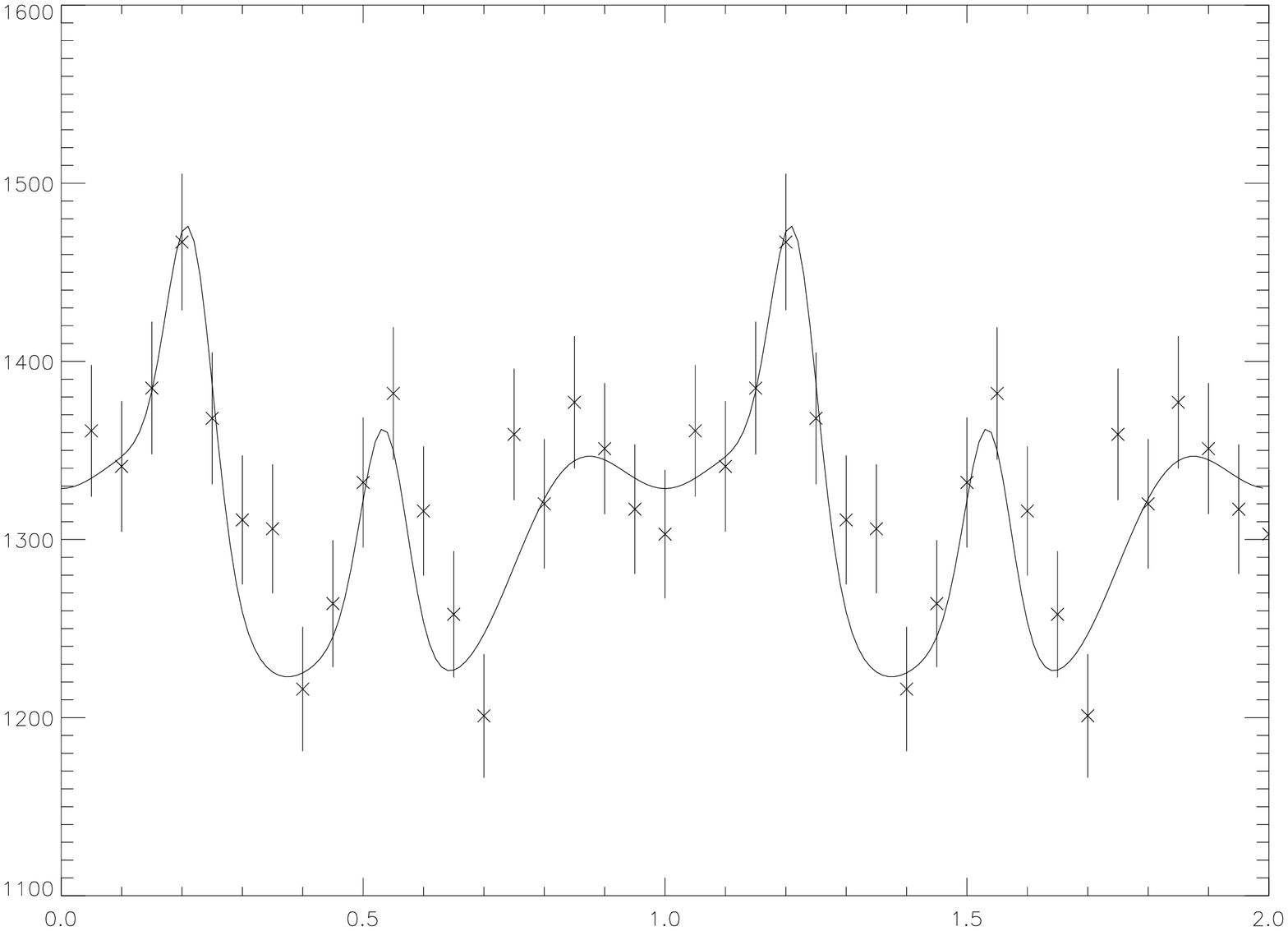,angle=90,width=3.5in}
\begin{minipage}[h]{3.5in}
{\small
Figure 3:  Post-glitch-12 light curve. October 96 data are
compared with the pre-glitch empirical fit of Figure 2.  Total
observation time was 43 ks.  This
observation started 15 days after glitch 12 and is our best chance,
in these data, to observe any post-glitch change in intensity or pulsed
waveform.  The post-glitch waveform is, within observational limits,
the same as the average pre-glitch waveform.
} 
\end{minipage}
\end{center}

Figure 3 shows the post-glitch October '96 data alone.  This
observation started 15 days after glitch 12 and is our best chance,
in these data, to observe any post-glitch change in intensity or pulsed
waveform.  The post-glitch waveform is, within observational limits,
the same as the average pre-glitch waveform.  The October '94
observation started 59 days after glitch 11 (which was small) and we
have included it in the sum of pre-glitch observations. \\

An independent technique was used to confirm that pulsed emission was
present.  The Princeton radio ephemeris was used to calculate the
pulsar period at the midpoint of each HRI data set.  Data were folded
at this period and light curves similar to those illustrated were
obtained.

\begin{center}
\section{INTERPRETATION OF PULSED SIGNAL}
\end{center}

We assume that the pulse has two components:  a broad
pulse and two narrow peaks.  There is also a
large non-varying (probably thermal) component. \\

We used a $10^{\prime \prime}$ radius circular region to extract the
pulsar data.  The pulsar appeared slightly elongated in all
observations because the ``wobble''  in the ROSAT pointing was not
completely removed.  This circle was large enough to encompass 85\% of
the events in the ROSAT point spread function.  Table 3 lists the
observed components of the HRI signal.  Uncertainty in the total rate
is due to counting statistics.  An order of magnitude larger 
uncertainty in the signal from the
pulsar is due to possible error in subtraction of a contribution
from the surrounding diffuse nebula. (Harnden \etal 1985) \\

To compare pre and post-glitch data we made an empirical fit to the
data shown in Figure 2.  This fit is indicated by the solid
line.  Because it appears more square than a pure sine wave, the
broad pulse form assumed is $\cos 2\pi\phi - (1/3) \cos 6\pi \phi$.
The two narrow peaks are assumed to be Gaussian with identical widths.
The fit to Figure 2 is  $I = 0.573 + 0.029 [\cos
2\pi\phi - 1/3 
\cos 6 \pi \phi] + 0.071 e^{- \frac{(\phi - .214)^{2}}{2(.038)^{2}}} +
0.056 e^{- \frac{(\phi - .535)^{2}}{2(.038)^{2}}}$. 
$\phi$ is phase (0-1) and the units of I are
ROSAT HRI counts s$^{-1}$.  The two narrow
peaks have FWHM of 0.09 in phase and are separated by 
0.32 in phase.  As summarized in table 3, after subtraction of a
nebular component, 12\% of the pulsar emission is pulsed with $\approx
8$\% in the broad pulse and $\approx 4$\% in the two narrow peaks. \\

The 1991 PSPC data were re-analyzed using the same techniques as for
the HRI data.  A 15$^{\prime \prime}$ radius region was used to
extract the pulsar signal and data were again folded at the
radio-determined pulsar frequency.  The April 1991 PSPC light
curve is compatible with 
the fit to the HRI data.  Epoch folding of the December 1991 data
produced no clean indication of the pulsar which probably indicates 
a timing problem in our December
1991 data. \\

\section{THE ABSOLUTE RATE}

The countrate of the pulsar (and immediate surroundings) was monitored
by extracting events from a circle 15$^{\prime \prime}$ in radius,
centered on the pulsar.  To minimize the effect of variations in
detector efficiency and background, the rate from the surrounding
annulus with 
inner and outer radii of 15$^{\prime \prime}$ and 80$^{\prime \prime}$
was also monitored and used for normalization.  This annulus
contains a relatively bright region of diffuse emission surrounding
the pulsar 
which should not vary on the time scale of this observation.  The
central region should contain 87\% of the pulsar emission (from the
ROSAT point response function in David et al
1998) and is large enough to allow for the observed attitude
smearing of a few arcseconds. \\

\begin{center}
\psfig{figure=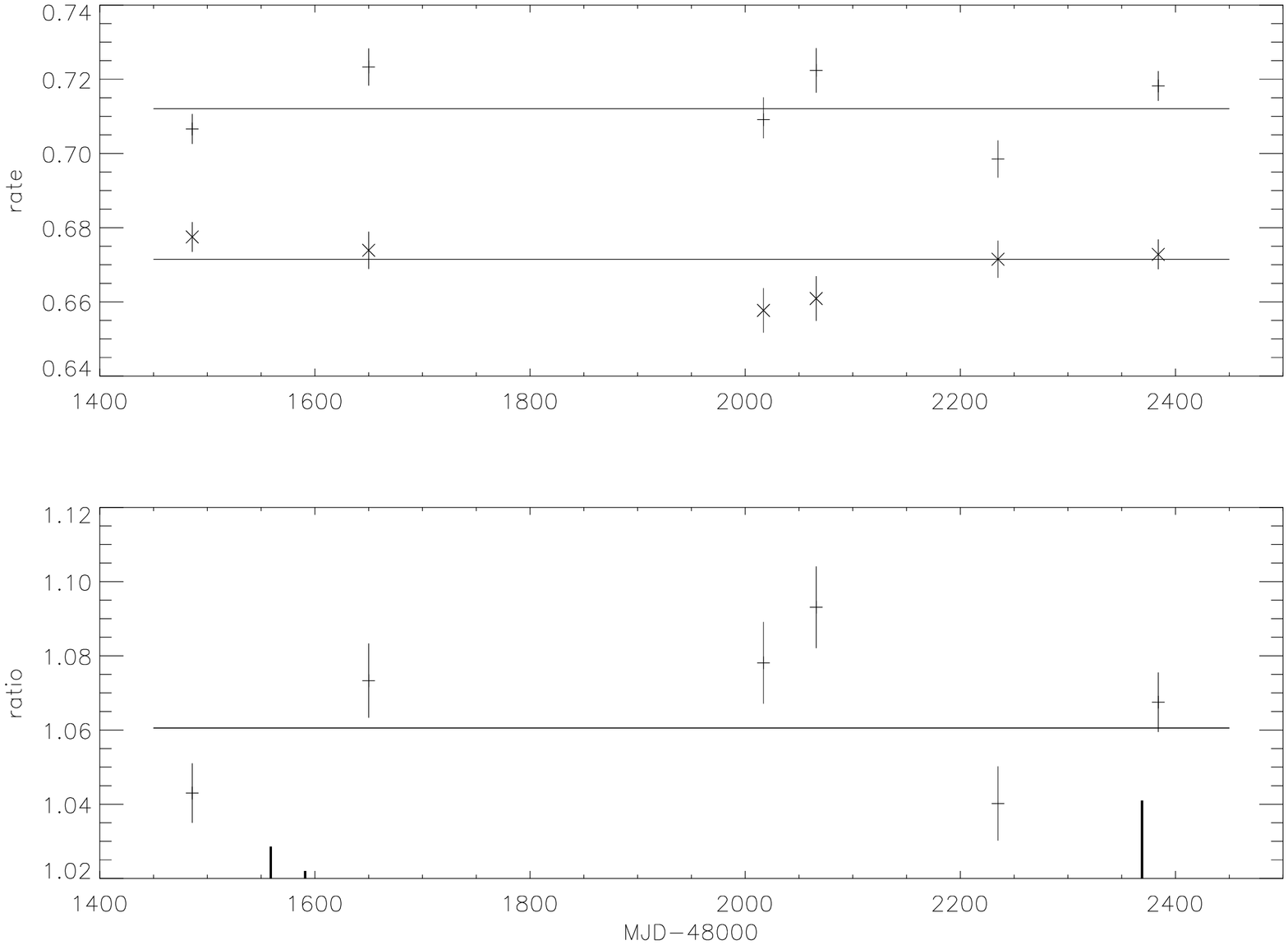,angle=90,width=3.5in}
\begin{minipage}[h]{3.5in}
{\small
Figure 4:  Pulsar brightness vs time.  Count rate from the pulsar and
immediate surroundings for 
the 6 HRI observations.  Top curve ($+$) is the rate taken from a
small region around the pulsar.  Middle curve (X) is from a
surrounding annulus and includes diffuse emission only.  Bottom curve
($+$) is the ratio of the two.  Horizontal lines are averages weighted
by exposure and glitches are indicated as bold vertical lines on the
time axis.  Error bars show $1\sigma$ counting statistics} 
\end{minipage}
\end{center}

These data are given in table 4 and plotted in Figure 4.  Error bars
show $1\sigma$ counting statistics only which are $\approx \pm 1$ \%.
No significant variation in the pulsar flux is seen, either in the
countrate or normalized countrate.  We note that there is a 3\%
increase in rate after both glitches (or rather a 3\% decrease in rate
before the glitches) and take this as an upper limit to an observed
effect.  A 3.0\% change in the rate shown in Figure 4 requires a 6.4\%
change in pulsar flux which, if attributed to black body emission, would
correspond to a temperature increase of 1.6\%.  We also note that the
accuracy inferred from Figure 4 is pushing the limit of the ROSAT
HRI.  Periodic calibrations with the supernova remnant N132D give apparent
efficiency variations of up to $\pm 5$\% for unknown reasons
(David \etal 1998).  Our
normalized rates should avoid this as well as any efficiency change
which might be associated with a detector gain (high voltage step)
change on June 21, 1994 (David \etal 1998), between the first and 
second HRI observations. \\

\begin{center}
\section{A SIMPLE ANALYSIS}
\end{center}

It is natural to assume that the narrow pulses are of non-thermal
origin.  The non-thermal gamma-ray pulses are both sharp
and double and the phase separation of the two X-ray pulses is
intermediate between the phase separation of the two optical and that
of the two gamma-ray pulses (Ramanamurthy, 1994). \\

The broad pulse could also be non-thermal.  The young pulsar,
B0540-69 exhibits a single broad pulse at both X-ray and optical
wavelengths (Deeter et al., 1999, Gouiffes et al., 1992, 
Manchester and Peterson, 1989).  The young pulsar, B1509-58, also
shows a single broad peak in the X-ray band with non-thermal spectrum
(Kawai et al 1993).  So it is interesting that the
Vela pulse waveform shows both narrow Crab-like pulses 
and the broad pulse associated with two other young pulsars. 
A thermal origin, however, for either broad or narrow Vela-Pulsar pulses is not
ruled out by these data.\\

Assuming the non-varying component to be thermal, we can derive 
the temperature and luminosity.  (We note that the validity of a
black-body approximation to the spectrum is untested and that the
derived temperature may not be physical.  Nevertheless, the result is
useful.) Using a distance of 500 pc, a black
body spectrum, a neutron star radiation  
radius of 10 km, and an interstellar column density of $5\times
10^{19}$ atoms/cm$^{2}$, the PIMMS program was used to calculate
the luminosity and surface temperature, $T$, from the ROSAT HRI count
rate.  Results are listed in Table 5.  Because the luminosity varies
as $T^{4}$, the 20\% uncertainty in count rate of the non-varying
component, if thermal, corresponds to a 5\% uncertainty in temperature.  A
more realistic range for parameters can be derived by using
extremes of the generally accepted ranges for distance (400-500 pc),
neutron star radius (7-15 km), and interstellar column ($N_H = 0.2-2\times
10^{20}$).  Possible values for $T$ lie in the range $5-11\times 10^{5}$
K with corresponding bolometric luminosities of $2-8\times 10^{32}$
erg/s. The most likely fit to our data yields a
neutron star temperature of $8.5\times 10^{5}$ K and bolometric
luminosity of $4\times 10^{32}$ erg/s.

\begin{center}
\section{APPLICATION TO GLITCH PHYSICS}
\end{center}

This null result can limit some values of parameters which describe neutron 
star structure. The parameters are largely unknown, are not independent, 
and can be linked through models. Theoretical models start with a model 
neutron star (a choice of mass and equation of state EOS). 
The remaining parameters include the initial (preglitch) temperature 
distribution, the amount of energy $\Delta$E dissipated in the glitch and the 
location, or density $\rho$, where the energy dissipation takes place. 
This heat, released promptly during the glitch, will diffuse out to the 
neutron star surface and inward, into the neutron star core. To calculate the 
postglitch thermal signal on the neutron star surface, models of the heat 
capacity and thermal conductivity, as well as neutrino emissivities from the 
interior are required. At the age of the Vela pulsar, the interior of the 
neutron star is expected to be isothermal with a core temperature extending 
through the inner crust. The associated surface temperature is determined by 
conductivities in the outer crust (Gudmundsson, Pethick and Epstein 1981). 
Calculations of the expected postglitch thermal signal on the surface 
have been reported by Eichler and Cheng (1989), van Riper, Epstein and 
Miller (1991), by Chong and Cheng (1994) and by Hirano et al (1996) for 
various combinations of the parameters. The results are qualitatively similar. 
All calculations are quite sensitive to the initial value of 
the surface temperature. The fractional change in 
temperature is higher for lower initial temperatures. For a conservative 
evaluation of our null results, in terms of which model parameters can be 
ruled out, we take the initial surface temperature to be the highest value, 
covered in the model calculations, 10$^6$ K. This is also the appropriate 
choice of initial surface temperature for the Vela pulsar. 
The ROSAT measurements of \"{O}gelman, Finley and Zimmermann (1993) yield surface 
temperatures of 1.5-1.6 $\times$ 10$^6$ K from black body fits to the point 
source and to the pulsed signal. The model-dependent analysis of the
previous section yields a temperature of 0.85 $\times$ 10$^6$ K.  
The actual value of the surface 
temperature may be somewhat different but of the same order when neutron star 
atmosphere models are used. \\

Van Riper et al (1991) present their results in terms of t$_{peak}$, the time 
at which the glitch-induced 
temperature signal on the surface is maximum; the width
$\Delta$t 
of this signal and the maximum fractional change in the surface temperature. 
Contours of t$_{peak}$ and $\Delta$t for which the postglitch temperature 
enhancement signal would fall within the range of our observations, 
15-22 days after the glitch, correspond roughly to neutron star radii 
R $<$ 13 Km. Thus  for T$_{s,0}$ = 10$^6$ K our upper limits test the models 
only for stars with R $<$ 13 Km. The radius of a typical 1.4 M$_{\odot}$ neutron 
star is 13.7 Km in the soft Baym, Pethick and Sutherland (1971) 
equation of state 
and 13.7 Km for the medium equation of state of Friedman and
Pandharipande (1981). 
For a stiff equation of state like the Pandharipande, Pines and Smith
(1976) model 
the radius of a 1.4 M$_{\odot}$ star is 18.6 Km, so our results do not test the 
models if the neutron star EOS is actually stiff. If the energy 
release $\Delta$E in the glitch was 10$^{42}$ ergs, the thermal signal would 
violate our upper bound if the glitch dissipated energy at 
densities $\rho \sim 10^{13}$ gm cm$^{-3}$ or less so it is likely that 
$\Delta$E $<$ 10$^{42}$ ergs, conditional on the $\rho$ values. 
$\Delta$E = 10$^{43}$ ergs can be ruled out altogether, for no matter at what 
$\rho$ the glitch energy was dissipated, the signal on the surface would 
violate our upper bounds.   \\

Chong and Cheng (1994) plot their results for 1.4 M$_{\odot}$ neutron stars of 
different EOS, initial surface temperatures and 
$\Delta$E , taking the density at which  the glitch energy is dissipated to be 
$\rho$ = 10$^{13}$ gm cm$^{-3}$ for all cases. Taking their model for a soft 
BPS star with the core temperature T$_c$ = 10$^8$ K, corresponding to an 
initial surface temperature of 1.27 $\times$ 10$^6$ K we find that 
$\Delta$E = 10$^{43}$ ergs is ruled out by our results. For the moderately 
stiff UT equation of state (Wiringa \& Fiks 1988) with T$_c$ = 10$^8$ K and 
T$_{s,0}$ = 10$^6$ K, the model prediction for the temperature enhancement at 
the times of our observations lie below the upper limits, so even 
$\Delta$E = 10$^{43}$ ergs is not ruled out. For the stiff PPS star model, 
with T$_c$ = 10$^8$ K and T$_{s,0}$ = 8 $\times$ 10$^5$ K, $\Delta$E = 
10$^{43}$ ergs is ruled out by our upper limits. \\

Hirano et al (1997) present similar results by marking in the A$_T$ $\cong$ 
$\Delta$ T$_{s}$/T$_{s,0}$ vs t$_{peak}$ plane the maximum A$_T$ points 
for different 1.4 M$_{\odot}$ neutron star models and different densities $\rho$ 
of energy release. Results are given for various values of the initial surface 
temperature T$_{s,0}$ and energy release $\Delta$E. To check 
if the predicted signals extend into our range of observation 
times, we utilized the temperature 
evolution for one-dimensional heat conduction, Eq.(4) of Hirano \etal which 
they quote as representative of the temperature evolution in  
their detailed numerical results. At T$_{s,0}$ = 10$^6$ K, 
$\Delta$E = 10$^{43}$ ergs is ruled out if the EOS is soft 
(the Baym, Pethick, Sutherland (1971) model).  For a moderately stiff FP star, 
our results rule out $\Delta$E = 10$^{43}$ ergs if the energy was released at 
$\rho$ $<$ 3 $\times$ 10$^{13}$ gm cm$^{-3}$. \\  

In addition to these model calculations, all of which assume that the glitch 
induced energy dissipation takes place in a spherically symmetric shell, a 
recent paper by Cheng, Li and Suen (1998) presents model calculations of the 
non-spherically symmetric case. These authors calculate the thermal signal 
on the neutron star surface for the case of glitch induced energy dissipation 
at densities 3 $\times$ 10$^{12}$ gm cm$^{-3}$ $<$ $\rho$ $<$ 3 $\times$
10$^{13}$ gm 
cm$^{-3}$, and within a 2 deg $\times$ 2 deg solid angle. For $\Delta$E = 
10$^{42}$ ergs a hot spot is found to emerge on the surface, keeping to 
the same solid angle range and peaking at 
274 days after the glitch. During our observations, 15-22 days after the 
glitch, a 50 percent enhancement in temperature is predicted in the hot spot. 
The contribution to the total luminosity of this hot spot remains well below 
our upper limits. The absence of significant changes in the pulse shape of the 
Vela pulsar before and after a glitch does not lead to a test of the 
models. The hot spot is at the same angular position as the region of energy 
dissipation in the glitch. For many values of the angles from the rotation 
axis and from our line of sight, there would be no modulation in the observed 
signal. Furthermore, the energy dissipation might take place in an equatorial 
belt around the rotation axis, as is likely if the glitch is due to vortex 
line unpinning (a rotational instability in the neutron star crust superfluid).\\

Returning to the spherically symmetric models we note that the differences 
between the model predictions are not important for purposes of comparison 
with our current upper limits. (These differences are likely to be due to 
different surface layer compositions in the models in terms of the A and Z of 
the equilibrium nuclear species, effecting the thermal conductivities 
(Hirano et al 1997)). Within the uncertainties indicated by these 
differences, and the large number of unknown parameters, we conclude that our 
upper bounds on a thermal signal rule out glitches with energy release, 
$\Delta$E, greater than 10$^{43}$ ergs. \\ 
 
For superfluid unpinning models of the glitches the energy dissipated is 
given by: 

\begin{equation}
\Delta E = I_p \delta\Omega \omega = I \Delta\Omega \omega.
\end{equation}

In these models the glitch reflects a transfer of angular momentum, J= 
$I_p \delta\Omega = I \Delta\Omega$, from a component of the neutron star, 
taken to be the pinned superfluid in the inner crust, to the rest of the star 
including the observed crust. I$_p$ is the moment of inertia of the pinned 
superfluid and $\delta\Omega$ the decrease in its rotation rate during 
the glitch. I is the moment of inertia of the observed 
crust and most of the neutron star, 
which is coupled rigidly to the crust. $\Delta\Omega$ is the observed  
jump in rotation frequency of the crust and $\omega$ is the difference in 
rotation rate between the superfluid and the normal crust. $\omega$ is related 
to the pinning forces. Our upper bound on $\Delta E$ implies a very loose 
upper bound $\omega$ $<$ 100 rad s$^{-1}$. Recent theoretical work 
(Pizzochero, Viverit \& Broglia 1997) gives pinning energies which imply
$\omega \sim$ 1 rad s$^{-1}$. The coupling between the pinned superfluid 
and the observed crust of the neutron star also entails a continuous 
rate of energy dissipation, realized though the thermal creep of vortices. In 
parallel to the glitch associated energy dissipation, this continuous rate 
is easily shown to be (Alpar et al 1984): 
\begin{equation}
\dot{E}_{diss} = I_p |\dot\Omega| \omega.
\end{equation}
This should supply the thermal luminosity of an old pulsar. 
Observational upper bounds from the thermal 
luminosities of older pulsars (Alpar et al 1987, Yancopoulos, 
Hamilton \& Helfand 1994) yield $\omega$ $<$ 1 rad s$^{-1}$. \\

While crustquake models fail to explain the magnitude and rate of glitches 
of the Vela pulsar, crustquakes may trigger the vortex unpinning and
angular momentum transfer that show up as glitches. The crust cracking
that triggers a 
glitch (Ruderman 1991) entails an energy dissipation 

\begin{equation}
\Delta E = \mu \theta^2
\end{equation}

where $\mu$ denotes the elastic energy modulus and $\theta$ is the critical 
strain angle for the breaking of the lattice. For a volume of linear 
dimensions 10$^5$ cm $\mu$ = 10$^{44}$ ergs. Our upper limit of 
$\Delta E$= 10$^{43}$ ergs then implies that $\theta$ $<$ 10$^{-1}$ rad s$^{-1}$. 
While terrestrial solids have $\theta \sim 10^{-4}$ or less, there have been 
speculations that the neutron star crust will have $\theta$ $>$ 10$^{-2}$, 
a much larger strength resulting from its unscreened Coulomb interactions. 
Thus, observations that are sensitive to implied glitch energy 
dissipation rates $\Delta E$ = 10$^{41}$ ergs will be critical for the
crust-breaking models of glitches as well as for the
superfluid-unpinning model. \\ 

While the present upper limit does not give any stringent constraints on the 
glitch related energy release, a future detection could yield important clues 
into the neutron star EOS, since in all calculations, different EOS yield 
different timescales for the postglitch thermal signal.  Post-glitch
RXTE-PCA observations of the Vela pulsar, at one, four and 
nine days, and at about three months following the October 1996 glitch 
do not constrain models of glitch related energy dissipation (G\"{u}rkan et 
al. 1999).
New instruments should have the capability to detect temperature rises from the 
Vela pulsar at the corresponding levels of sensitivity. The Chandra HRC 
detector is 3.4 times as sensitive as the ROSAT HRI and can thus collect data 
equivalent to that shown in Figure 2 in 36 ks. Two such observations, one day 
and one week after a glitch, should measure an increase in surface luminosity 
if the energy release is 10$^{41}$ ergs or greater. This time frame does not 
cover all of parameter space but is probably the best place to
search. \\

This work was supported by NASA through contract NAS8-39073 and grant
NAG5-6853. \\

{}

\newpage

\begin{center}

\begin{tabular}{llll} \hline 
\multicolumn{4}{c}{Table 1} \\ \hline
\multicolumn{1}{c}{~~~}&
\multicolumn{1}{c}{Rotation}&
\multicolumn{1}{c}{Frequency} \\
\multicolumn{1}{c}{Date}&
\multicolumn{1}{c}{Frequency}&
\multicolumn{1}{c}{Derivative}&
\multicolumn{1}{c}{Epoch}  \\
\multicolumn{1}{c}{(UT)}&
\multicolumn{1}{c}{(Hz)}&
\multicolumn{1}{c}{$10^{-12}$ (Hz s$^{-1}$)}&
\multicolumn{1}{c}{(MJD)} \\ \hline
14-26 May 94&11.197 366 280 36&-15.5881&49492.274659780 \\
25-28 Oct 94&11.197 160 398 75&-15.6554&49653.5062011504 \\
27 Oct -2 Nov 95&11.196 668 608 990&-15.5921&50017.020877910 \\
15-16 Dec 95&11.196 603 154 274&-15.5858&50066.015970352 \\
1-3 Jun 96&11.196 376 285 216&-15.5777&50236.139720991 \\
28 Oct - 4 Nov 96&11.196 198 755 884&-15.5693&50385.205072337 \\
\end{tabular}

\vspace*{0.5in}

\begin{tabular}{llllll} \hline
\multicolumn{6}{c}{Table 2} \\ \hline
\multicolumn{1}{c}{~~~~}&
\multicolumn{1}{c}{~~~~}&
\multicolumn{1}{c}{Exposure}&
\multicolumn{1}{c}{X-ray}&
\multicolumn{1}{c}{Period} \\
\multicolumn{1}{c}{Date}&
\multicolumn{1}{c}{Instrument} &
\multicolumn{1}{c}{Live time} &
\multicolumn{1}{c}{Period}&
\multicolumn{1}{c}{Derivative}&
\multicolumn{1}{c}{Epoch} \\
\multicolumn{1}{c}{(UT)}&
\multicolumn{1}{c}{~~~~}&
\multicolumn{1}{c}{(s)}&
\multicolumn{1}{c}{(s)}&
\multicolumn{1}{c}{$10^{-13}$(s s$^{-1}$)}&
\multicolumn{1}{c}{(MJD)} \\ \hline
14-26 May 94&HRI&62090&0.089 306 649 49&1.2433&49486.169164815 \\
25-28 Oct 94&HRI&31545&0.089 308 327 69&1.2486&49650.775217647 \\
27 Oct - 2 Nov 95&HRI&19409&0.089 312 279 83&1.2437&50017.896366423 \\
15-16 Dec 95&HRI&20380&0.089 312 801 95&1.2432&50066.494671817 \\
1-3 Jun 96&HRI&28834&0.089 314 611 67&1.2427&50235.015672685 \\
28 Oct - 4 Nov 96&HRI&43194&0.089 316 027 86&1.2520&50384.625402581 \\
\end{tabular}

\vspace*{0.5in}

\begin{tabular}{ll} \hline
\multicolumn{2}{c}{Table 3} \\ \hline
Total rate within $10^{\prime \prime}$ radius circle (s$^{-1}$)&0.585
$\pm$ .005 \\
non X-ray and diffuse background&0.0007 $\pm$ .0001\\
contribution from surrounding nebula&0.258 $\pm$ .066 \\
total signal from Vela PSR&0.326 $\pm$ .066 \\
2 narrow pulses&0.0122 $\pm$ .003 \\
broad-pulse&0.0277 $\pm$ .007 \\
steady (thermal?)&0.286 $\pm$ .066 \\ \hline
\end{tabular}

\vspace*{0.5in}

\begin{tabular}{lll} \hline
\multicolumn{3}{c}{Table 4} \\ \hline
\multicolumn{1}{l}{Date}&
\multicolumn{1}{l}{Average Rate}&
\multicolumn{1}{c}{Normalized Rate} \\ \hline
May 94&0.707$\pm$.004&1.043$\pm$.008 \\
Oct 94&0.723$\pm$.005&1.073$\pm$.010 \\
Oct 95&0.709$\pm$.006&1.078$\pm$.011 \\
Dec 95&0.722$\pm$.006&1.093$\pm$.011 \\
Jun 96&0.699$\pm$.005&1.040$\pm$.010 \\
Oct 96&0.718$\pm$.004&1.067$\pm$.008 \\
\end{tabular}

\vspace*{0.5in}

\begin{tabular}{cccccc} \hline
\multicolumn{6}{c}{Table 5} \\ \hline
\multicolumn{1}{c}{~~~~~}&
\multicolumn{1}{c}{ISM}&
\multicolumn{1}{c}{Neutron Star}&
\multicolumn{1}{c}{~~~~~}&
\multicolumn{1}{c}{~~~~~}&
\multicolumn{1}{c}{~~~~~} \\
\multicolumn{1}{c}{Distance}&
\multicolumn{1}{c}{Column}&
\multicolumn{1}{c}{Radius}&
\multicolumn{1}{c}{kT}&
\multicolumn{1}{c}{T}&
\multicolumn{1}{c}{L$_{bol}$} \\
\multicolumn{1}{c}{(pc)}&
\multicolumn{1}{c}{($10^{20}\mbox{cm}^{-2}$)}&
\multicolumn{1}{c}{(km)}&
\multicolumn{1}{c}{(keV)}&
\multicolumn{1}{c}{($10^{5}$K)}&
\multicolumn{1}{c}{($10^{32}\mbox{erg/s}$)} \\ \hline
500&0.5&7&.082&9.5&3.7 \\
500&0.5&10&.070&8.5&3.8 \\
500&0.5&15&.055&6.4&4.1 \\ \hline
500&0.2&10&.066&7.7&3.0 \\
500&2&10&.082&9.5&7.0 \\ \hline
extremes&&&&& \\ \hline
500&2&7&.095&11&8.2 \\
400&0.2&15&.047&5.5&1.9 \\ \hline
\end{tabular}
\end{center}

\end{document}